\documentclass{article}

\usepackage{amsmath, amsfonts}

\def\eqref#1{equation~\ref{#1}}

\def\rmX{{\mathbf{X}}}
\def\rmY{{\mathbf{Y}}}

\def\vq{{\bm{q}}}

\def\vv{{\bm{v}}}

\def\vx{{\bm{x}}}
\def\vy{{\bm{y}}}

\DeclareMathAlphabet{\mathsfit}{\encodingdefault}{\sfdefault}{m}{sl}
\SetMathAlphabet{\mathsfit}{bold}{\encodingdefault}{\sfdefault}{bx}{n}

\usepackage[T1]{fontenc} 
\usepackage[utf8]{inputenc} 
\usepackage{ismir,amsmath,cite,url}
\usepackage{graphicx}
\usepackage{subcaption}
\usepackage{microtype}
\usepackage{paralist}
\usepackage{color}
\usepackage{units}
\usepackage{bm}
\usepackage{flexisym}

\usepackage[bookmarks=false,hidelinks]{hyperref}

\title{An Attention Mechanism for Musical Instrument Recognition}

\multauthor
{Siddharth Gururani$^1$ \hspace{1cm} Mohit Sharma$^2$ \hspace{1cm} Alexander Lerch$^1$} { \\
 $^1$ Center for Music Technology, Georgia Institute of Technology, USA\\
$^2$ School of Interactive Computing, Georgia Institute of Technology, USA\\
{\tt\small \{siddgururani, mohit.sharma, alexander.lerch\}@gatech.edu}
}
\def\authorname{Siddharth Gururani, Mohit Sharma, Alexander Lerch}

\begin{document}

\maketitle
\begin{abstract}
  While the automatic recognition of musical instruments has seen significant progress, the task is still considered hard for music featuring 
  multiple instruments as opposed to single instrument recordings. Datasets for polyphonic instrument recognition can be 
  categorized into roughly two categories. Some, such as MedleyDB, have strong per-frame instrument activity annotations but 
  are usually small in size. Other, larger datasets such as OpenMIC only have weak labels, i.e., instrument 
  presence or absence is annotated only for long snippets of a song. We explore an attention mechanism for 
  handling weakly labeled data for multi-label instrument recognition. Attention has been found to perform well 
  for other tasks with weakly labeled data. We compare the proposed attention model to multiple models which include 
  a baseline binary relevance random forest, recurrent neural network, and fully connected neural networks. Our 
  results show that incorporating attention leads to an overall improvement in classification accuracy metrics across 
  all 20 instruments in the OpenMIC dataset. We find that attention enables models to focus on (or `attend to') specific 
  time segments in the audio relevant to each instrument label leading to interpretable results.
\end{abstract}
\section{Introduction}\label{sec:introduction}

Musical instruments, both acoustic and electronic, are necessary tools to create music. Most musical pieces comprise of a 
combination of multiple musical instruments resulting in a mixture with unique timbre characteristics.
Humans are fairly adept at recognizing musical instruments in the music they hear. Recognizing instruments automatically, 
however, is still an active area of research in the field of Music Information Retrieval (MIR). 
Instrument recognition in isolated note or single instrument recordings has achieved a fair amount of success \cite{han2016sparse,lostanlenExtendedPlayingTechniques2018}.
Recognizing instruments in music with multiple simultaneously playing instruments, however, is still a hard problem. The task is difficult because of \begin{inparaenum}[(i)]
  \item the superposition (in both time and frequency) of multiple sources/instruments, 
  \item the large variation of timbre within one instrument, and
  \item the lack of annotated data for supervised learning algorithms.
\end{inparaenum}

Identifying music in audio recordings is helpful for general retrieval systems by allowing users to search for music with specific instrumentation \cite{takahashiInstrudiveMusicVisualization2018}.
Instrument recognition can also be helpful for other MIR tasks. For example, instrument tags may be vital for music recommendation systems to model users' affinity towards certain instruments, genre recognition systems could also improve with genre-dependent instrument information. Building models conditioned on a reliable detection of instrumentation could also lead to improvements for tasks such as automatic music transcription, source separation, and playing technique detection.

As mentioned above, one of the challenges in MIR in general, and in instrument recognition in particular, 
is the lack of large-scale annotated or labeled data for supervised machine learning algorithms \cite{wu_labeled_2018, humphrey2018openmic}. 
Datasets for instrument recognition in polyphonic music can broadly be divided into strongly and weakly labeled.
%
%
A weakly labeled dataset (WLD) contains clips that may be several seconds long and have labels for one or more instruments for their entirety without annotating the exact onset and offset times of the instruments. A strongly labeled dataset (SLD), however, contains audio with fine-grained labels of instrument activity. 
WLDs are easier to annotate compared to SLDs and therefore scale better. Even though SLDs enable strong supervision of learning algorithms, the smaller size may lead to poor performance of deep learning methods. 
WLDs, however, have the disadvantage that an instrument may be marked positive even if the instrument is active for a very short duration of the entire clip. 
This makes it challenging to train models with WLDs.

Models for recognition in weakly labeled data may benefit from inferring the specific location in time of the instrument to be recognized. We formulate the polyphonic instrument recognition task as a multi-instance multi-label (MIML) problem, where each weakly labeled example is a collection of short-time instances, each with a contribution towards the labels assigned to the example.
Toward that end, we apply an attention mechanism to aggregate the predictions for each short-time instance and compare this approach to other models which include binary-relevance random forests, fully connected networks, and recurrent neural networks. We hypothesize that the ability of the attention model to weigh relevant and suppress irrelevant predictions for each instrument leads to better classification accuracy. We visualize the attention weights and find that the model is able to mostly localize the instruments, thereby enhancing the interpretability of the classifier.

The next section reviews literature in instrument recognition and audio tagging or classification. Sect.~\ref{sec:data_challenges} discusses various datasets for instrument recognition and the challenges associated. Next, Sect.~\ref{sec:method} formulates the problem and describes the model. Sect.~\ref{sec:eval} specifies the various experiments and the evaluation metrics to measure performance. We report the results of the experiments and discuss them in Sect.~\ref{sec:discussion}. Finally, in Sect.~\ref{sec:conclusion} we conclude the paper suggesting future directions for research.

\section{Related work}\label{sec:related}


\subsection{Musical Instrument Recognition}\label{sec:related_poly}

Instrument recognition in audio containing a single instrument can refer to both recognition from isolated notes or recognition from solo recordings of pieces. 
We refer to \cite{lostanlenExtendedPlayingTechniques2018, herrera2003automatic} for a review of literature in single instrument and monophonic instrument recognition.

Current research has focused on instrument recognition in polyphonic and multi-instrument recordings. While traditional approaches extract features followed by classification algorithms were previously prevalent \cite{kitahara2007instrument, fuhrmann2012automatic}, deep neural networks have dominated recent work in this field.
Han et al.\ \cite{han2017deep} applied Convolutional Neural Networks (CNNs) to the task of predominant instrument recognition on the IRMAS dataset \cite{bosch2012comparison} and outperformed various feature-based techniques. 
Li et al.\ \cite{li_automatic_2015} proposed to learn features from raw audio using CNNs for instrument recognition using the MedleyDB dataset \cite{bittner2014medleydb}. 
Gururani et al.\ \cite{gururani_instrument_2018} compared various neural network architectures for instrument activity detection using two multi-track datasets containing fine-grained instrument activity annotations: \begin{inparaitem}[]
  \item MedleyDB and
  \item Mixing Secrets \cite{gururani_mixing_2017}
\end{inparaitem}.
They found significant improvement of CNNs and Convolutional Recurrent Neural Networks (CRNNs) over fully connected networks and proposed a method for visualizing model confusion in a multi-label setting.
Hung et al.\ \cite{hung_frame_2018} utilized the fine-grained instrument activity as well as pitch annotations in the MusicNet dataset \cite{thickstun_learning_2017} and showed the benefits of pitch-conditioning on instrument recognition performance.
In follow-up research, Hung et al.\ \cite{hung_multitask_2018} proposed a multi-task learning approach for instrument recognition involving the prediction of pitch in addition to instrumentation. They released a synthetic, large-scale, and strongly-labeled dataset generated from MIDI files for evaluation and found that multi-task learning outperforms their previous approach of using pitch features as additional inputs.


\subsection{Audio event detection, tagging and classification}\label{sec:related_audio}

%

The task of audio or sound event classification shares many commonalities with instrument recognition.
Both tasks aim to identify a time-variant sound source in a mixture of multiple sound sources. 
A few key differences are that research in sound event classification typically focuses on uncorrelated sounds such as motor noise, car horns, baby cries, or dog barks, while musical audio is highly correlated. 
Additionally, music has a rich harmonic and temporal structure usually absent in audio captured from real world acoustic scenes.

For a historic review of work in sound event and audio classification, we refer readers to the survey article by Stowell et al.\ \cite{Stowell2015}. We focus on more recent literature involving deep neural network architectures~---which are now the standard approach---~as well as on methods that focus on addressing weak labels.

Hershey et al.\ \cite{hershey_CNN_2017} adapted deep CNN architectures from computer vision and found that they are effective for large-scale audio classification.
Cakir et al.\ \cite{cakir2017convolutional} researched the benefits of CRNNs for sound event detection over models comprising of only CNNs.
They found that the ability of RNNs to capture long-term temporal context helps improve performance against models only comprising CNNs. 
Adavanne et al.\ \cite{adavanne_sound_2017} proposed to use spatial features extracted from multi-channel audio as inputs for CRNN architectures. 
They found that presenting these features as separate layers to the model outperforms concatenation of these features at the input stage.


Learning from weakly labeled data has also been a focus in audio classification. Most works utilize the Multiple-Instance Learning (MIL) framework for the task, where each example is a labeled bag containing multiple instances whose labels are unknown. 
Kumar and Raj \cite{Kumar_AED_2016} utilized support vector machines and neural networks for solving the MIL problem. They train bag-level classifiers capable of predicting instances and are hence also useful for localization of sound events. 
Similarly, Kong et al.\ \cite{kong_audio_2018} proposed decision-level attention to solve the MIL problem for Audio Set \cite{gemmeke_Audio_2017} classification. 
Attention is applied to instance predictions to enable weighted aggregation for bag-level prediction. 
Kong et al.\ \cite{kong_weakly_2019} extended this and propose feature-level attention where instead of applying attention to the instance predictions, it is applied to the hidden layers of a neural network to construct a fixed-size embedding for the bag. 
Finally a fully connected network predicts the labels for the bag using the embedding vector.
McFee et al.\ \cite{mcfee_adaptive_2018} compared various methods for aggregating or pooling instance-level predictions. They developed an adaptive pooling operation capable of interpolating between common pooling operations such as mean-, max- or min-pooling.

\section{Data challenge}\label{sec:data_challenges}

In Sec.~\ref{sec:related_poly}, we introduced research on instrument recognition in polyphonic, multi-timbral music. One theme that emerges is that with almost every new publication, a new dataset is released by the authors in an effort to address issues with previous ones. While releasing new datasets is highly encouraged and vital for research in MIR in general, an uncoordinated effort leads to lack of uniformity in the datasets used.
In this section we briefly describe the common datasets for instrument recognition and identify the challenges associated with them. 

The IRMAS dataset \cite{bosch2012comparison} is a frequently used dataset for predominant instrument recognition. It consists of a separate training and testing set, each containing annotations for 11 predominant instruments.
The dataset consists of short excerpts~---\unit[3]{s} for training and variable length for testing---~of weakly labeled data.
One fundamental problem of the IRMAS annotations is that the training set lacks multi-label annotation; this can be problematic for a general use case as instrument co-occurrence is ignored. 

The MedleyDB \cite{bittner2014medleydb} and Mixing Secrets \cite{gururani_mixing_2017} datasets are both multi-track datasets. Due to the availability of instrument-specific stems, strong annotations of instrument activity are available. Thus, these two multi-track datasets provide all the necessary detailed annotations for instrument activity detection and have been used in \cite{gururani_instrument_2018,li_automatic_2015}. These datasets have two disadvantages when training models. First, with a few hundred distinct songs models trained with the data are hardly generalizable.
Second, the datasets are not well balanced in terms of either musical genre or instrumentation. However, this may not be a problem if the datasets were larger and the distribution represented the real-world.

Most of these problems were addressed with the release of the OpenMIC dataset \cite{humphrey2018openmic}. This dataset contains 20,000 \unit[10]{s} clips of audio from different songs across various genres. Each clip is annotated with the presence or absence of one or more of $20$ instrument labels. OpenMIC presents a larger sample size as well as a uniform distribution across instruments. It is, however, weakly labeled, i.e., each \unit[10]{s} clip has instrument presence or absence tags without specific onset and offset times.
Due to the nature of weak labels, models cannot be trained using fine-grained instrument activity annotation as done, e.g., in \cite{hung_frame_2018, gururani_instrument_2018}. Additionally, not all clips are labeled with all $20$ instruments, i.e., there are missing labels. This complicates the training procedure if models are to predict the presence/absence for all 20 instruments for an input audio clip.
Despite their drawbacks, creation of WLDs scales better since weak labels are cheaper to obtain; models capable of exploiting WLDs may thus be vital for the future development of instrument recognition.


\section{Method}\label{sec:method}

Before describing the model details, we provide a formalization of our approach to the instrument recognition problem in weakly labeled data.

\subsection{Pre-Processing}\label{sec:preprocess}
As mentioned in Sect.~\ref{sec:data_challenges}, the OpenMIC dataset consists of \unit[10]{s} audio clips, each labeled with the presence or absence of one or more of $20$ instrument labels. For each audio file in the dataset, the dataset creators also release features extracted from a pre-trained CNN, known as ``VGGish'' \cite{hershey_CNN_2017}. The VGGish model, based on the VGG architectures for object recognition \cite{simonyan2015very}, is trained for audio classification. 
The model produces a $128$-dimensional feature vector for \unit[0.96]{s} windows of audio with no overlap. The features are ZCA-whitened and quantized to $8$-bits. For a \unit[10]{s} audio file, we obtain a $10\times128$-dimensional matrix. We also normalize the 8-bit integers to a quantized range of $[0,1]$.

\subsection{Formulation}

\subsubsection{Multi-Instance Multi-Label Problem}
In the most general setting, instrument recognition can be framed as Multi-Instance Multi-Label (MIML) classification \cite{miml-scene.zhi.2007, miml-framework.zhi.2008, zha2008joint}.
Under this setting, we are given a training dataset $\{(\rmX_1, \rmY_1), \ldots (\rmX_m, \rmY_m)\}$ where $\rmX_i$ is a bag containing $r$ instances
$X_i = \{ \vx_{i, 1}, \ldots \vx_{i, r}\}$ and $\rmY_i = [\vy_{i,1}, \ldots, \vy_{i, L}] \in \{0,1\}^L$ is a label vector with $L$
labels with $\vy_{i,j} = 1$ if any of the instances in $\rmX_i$ contains label $j$. 
In the remainder of this section, we will drop the indices used to reference a specific data point and simply represent a sample from the 
dataset as $(\rmX, \rmY)$. In our case, a bag $\rmX$ refers to the $10\times128$-dimensional feature matrix representing one audio clip and each bag contains $10$ instances.
Our problem is also a Missing Label problem since for a sample $(\rmX, \rmY)$, not all $\vy_{j}$ are known or annotated (compare Sect.~\ref{sec:data_challenges}).

In our experiments, we assume that all labels can be independently predicted for each instance.
Under this assumption, the MIML problem decomposes into $L$ (20 for OpenMIC dataset) instantiations of Multi-Instance Learning (MIL) \cite{foulds2010review, zhou2002neural} problems, one for each label in the dataset.

Note that exploiting label-correlation in multi-label classification has shown to significantly improve the classification performance \cite{trohidis2008multi, zhang2010multi, qi2007correlative, qi2007correlative}. 
However, exploring ways to incorporate label-correlation for instrument recognition in the OpenMIC dataset has the additional challenge of missing and sparse labels \cite{bi2014multilabel}.
Also, as is prevalent in most MIL approaches \cite{foulds2010review}, we assume independence among different instances in a bag.
Neighboring instances in a bag representing a polyphonic music snippet will, however, likely have high correlation.
Relaxing the aforementioned assumptions about independence among labels, and instances in a bag is left for future work since in our current work, we focus on the impact of attention for aggregating instance-level predictions.

\begin{figure}
  \centering
  \includegraphics[width=.8\linewidth]{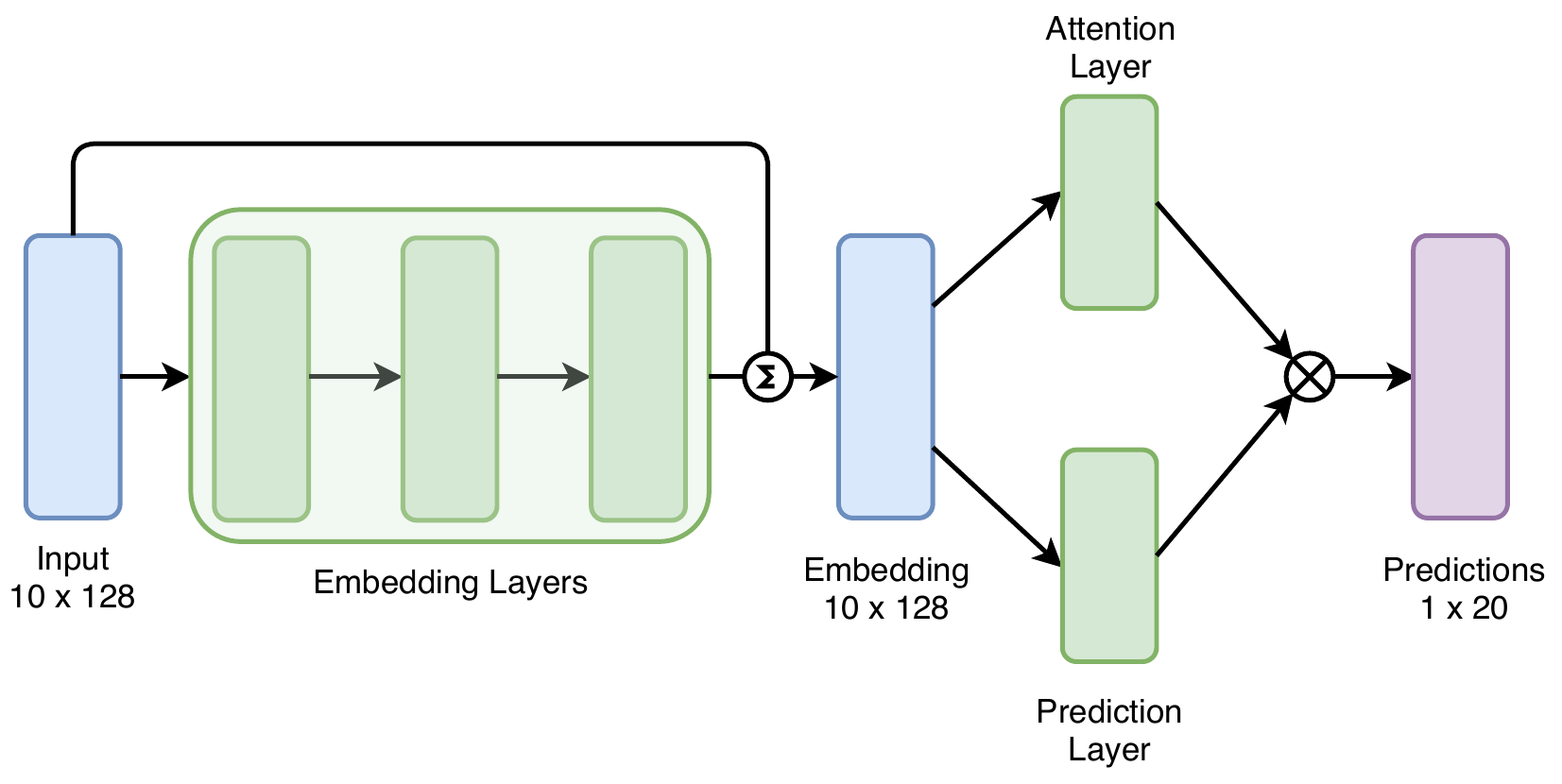}
  \caption{Model Architecture}
  \label{fig:fig1}
\end{figure}

\subsubsection{Multi-Instance Learning}
In the MIL setting, a bag label is produced through a score function $S(\rmX)$.
Under the assumption of independence among instances, $S(\rmX)$ admits a parametrization of the form
\begin{align}
  S(\rmX) = \mu \bigg( \; f(\vx) \; \bigg)
  \label{eq:score_bag}
\end{align}
where $f(.)$ is a score function for an instance $\vx$, and
$\mu(.)$ is a permutation-invariant aggregation operation for instance scores $f(\vx)$ \cite{deep-sets.zaheer.2017}.
This parameterization induces a natural approach to classify a bag of instances:
\begin{inparaenum} [(i)]
\item to produce scores for each instance in the bag using an instance-level scoring function $f(\vx)$, and
\item to aggregate the scores across different instances in the bag using the aggregation function $\mu(.)$.
\end{inparaenum}
In our approach, we use a classification function to produce instance-level scores $f(\vx)$, which are essentially the probabilities of a label being present for each instance.
The $max$ and $avg$ functions are two commonly used permutation-invariant operations to aggregate instance-level scores to bag-level scores.
McFee et al.\ found that \textit{learning} an aggregation operation, however, significantly improved performance over fixed predefined operations like 
$max$ and $avg$.
We choose to represent our aggregation operation $\mu(.)$ as a weighted sum of instance-level scores, i.e.,
\begin{align}
S(\rmX) = \sum_{\vx \in \rmX} w_{\vx} \; f(\vx)
\label{eq:bag_score}
\end{align}
where $w_{\vx}$ is a learnable weight for instance $\vx$.
Our choice of $f(.)$ and $\mu(.)$ has the two advantages that
\begin{inparaenum} [(i)]
\item the resulting $S(.)$ is the probability of a label being present in the bag and can be directly used to make a prediction and 
\item  the learned weights for each instance add interpretability to the MIL models by encoding beliefs placed by the MIL model on the score of each instance. 
\end{inparaenum}

\subsubsection{Attention Mechanism} 
The learnable aggregation operation is equivalent to attention.
Given a bag $\rmX = \{\vx_1, \ldots, \vx_r\}$ of $r$ instances, the instance level scoring function $f(.)$ produces a bag $\{f(\vx_1), \ldots, f(\vx_r)\}$ of instance scores.
The bag-level score $S(\rmX)$ is then computed using Eq.~(\ref{eq:bag_score}).

We further impose the restriction that instance weights $w_x$ should sum to 1, {i.e.}, $\sum_{\vx \in \rmX}{w_{\vx}} = 1$. 
This ensures that the aggregation operation is invariant to the size of the bag, thus allowing the model to work with sound clips of arbitrary length.
Furthermore, this normalization leads to a probabilistic interpretation of the instance weights which can then be
used to infer the relative contribution of each instance towards $S(\rmX)$.
For an instance $\vx \in \rmX$, the weight $w_{\vx}$ is thus parametrized as
\begin{align}
  w_{\vx} = \frac{\sigma\Big(\vv^\top h(\vx)\Big)}{\sum_{\vx\textprime \in \rmX} \sigma\Big( \vv^\top h(\vx\textprime)\Big)}
  \label{eq:attention}
\end{align}
where $h(\vx)$ is a learned embedding of the instance $\vx$, $\vv$ are the learned parameters of the attention layer, and $\sigma(.)$ is the \textit{sigmoid} non-linearity.

This corresponds to the attention mechanism traditionally used in sequence modeling \cite{bahdanau_neural_2015, vaswani_attention_2017}.
For example, Raffel and Ellis \cite{raffel_2015_feedforward} produced attention weights in a manner similar to Eq.~(\ref{eq:attention}) with the only difference being the use of \textit{softmax} operation to perform normalization of weights across the instances.

\subsection{Model Architecture}
Computing bag-level scores $S(.)$ involves computing instance-level scores $f(.)$ and aggregating the scores across instances using a learned set-operator $\mu(.)$ which performs weighted averaging with the weights computed with Eq.~(\ref{eq:attention}). 
For our experiments, we represent the scores, both instance level $f(.)$ and bag-level $S(.)$, as the probability
estimate of the instance or bag being a positive sample for a given label.
We first pass each instance $\vx$ through an embedding network of three fully connected layers to project each instance to a suitable embedding space.
Next, instance-level scores $f(.)$ are computed from the output of embedding network with another fully connected layer.
Similarly, attention weights are computed by normalizing the outputs of a fully connected layer, the weights of
which correspond to parameters $\vv$ in Eq.~(\ref{eq:attention}). Note that the output dimension of these two parallel fully connected layers is equal to the number of labels, i.e., $20$. Figure~\ref{fig:fig1} illustrates the model architecture. In the embedding layer, the number of hidden units is $128$. We also found that adding a skip connection from the input to the final embedding stabilized the training across different random seeds. We use batch normalization, ReLU activations, and a dropout of $0.6$ after each embedding layer. The model has 55336 learnable parameters.

%

\subsection{Loss Function and Training Procedure}\label{sec:partial_bce}

Our model performs a multi-label classification over 20 labels given an input. 
However, as we point out earlier, the OpenMIC dataset does not contain all labels for each instance.
This leads to missing ground truth labels for training with loss functions such as binary cross-entropy (BCE).
To account for this, we utilize the partial binary cross-entropy ($\mathrm{BCE}_{p}$) loss function introduced for handling missing labels \cite{durand_learning_2019}:

\begin{align}
    \begin{split}
        \text{BCE}_{p}(\vy, \vq) &= \frac{g(p_y)}{L}\sum_{l \in L^o} \vy_l \log \vq  + (1-\vy_l)\log(1-\vq) \\
        g(p_y) &= \alpha p_y^\gamma + \beta
    \end{split}
    \label{eq_4}
\end{align}
Here $g(p_y)$ is a normalization function, $p_y$ is the proportion of observed labels for the current data point, $L$ is the total number of labels, $L^o$ is the list of observed labels for the input data, $\vy_l \in \{0,1\}$ is the ground truth (absent or present) for label $l$, and $\vq$ is the model's probability output for the label $l$ being present in the input data $\rmX$.
The hyperparameters in Eq.~(\ref{eq_4}) are $\alpha$, $\beta$, and $\gamma$. Note that in the absence of $g(p_y)$, data points with few observed labels will have a lower contribution in loss computation than those with several observed labels. This is undesirable behavior and the inclusion of a normalization factor, dependent on the proportion of observed labels, is important. Therefore, we set $\alpha$, $\beta$, and $\gamma$ to $1$, $0$, and $-1$, respectively. This normalizes the loss for a data point by the number of observed labels and is equivalent to only computing the loss for observed labels.

Finally, the Adam optimization algorithm \cite{kingma_adam_2015} is used for training with a batch size of $128$ and learning rate of $5e^{-4}$ for $250$ epochs. We checkpoint the model at the epoch with the best validation loss. 

\section{Evaluation}\label{sec:eval}

In this section we describe the experimental setup including the dataset, the baseline methods, and evaluation metrics.
\subsection{Dataset}

We use the OpenMIC dataset for the experiments in this paper. In addition to the audio and label annotations, the data repository contains pre-computed features extracted from the publicly available VGG-ish model for audio classification.
We utilize those features in our experiments to strictly focus on handling the weak labels and avoid further complexity by having to learn features from the raw data or spectrogram representations. 
Pilot experiments for feature learning showed that CNN architectures based on state-of-the-art instrument recognition models were unable to outperform the baseline model of $20$ instrument-wise random forest classifiers trained using the pre-computed features.
For reproducibility and comparability, we utilize the training and testing split released with the dataset. Additionally, we randomly sample and separate $15\%$ data from the training split to create a validation set. 

\subsection{Experiments}

We compare the attention model (ATT) with the following models:
\begin{compactenum}
  \item RF\_BR: This model is the baseline random forest model in \cite{humphrey2018openmic}. A binary-relevance transformation is applied to convert the multi-label classification task into $20$ independent binary classification tasks \cite{Zhang2018binary}.
  
  \item FC: A 3-layer fully connected network trained to predict the presence or absence of all instruments for a given data instance. Here, the input features of dimension $10\times 128$ are flattened into a single feature vector for classification. Dropout is used for regularization and the Leaky ReLU ($0.01$ slope) is used. The model has 986772 parameters.

  \item FC\_T: This model serves as an ablation study to observe the benefits of the attention mechanism. FC\_T uses the same embedding layer as ATT. However, the aggregation of predictions in time is simply performed with average-pooling. The model has 52116 parameters.

  \item RNN: A 3-layer bi-directional gated recurrent unit model with 64 hidden units per direction. The model processes the input features and produces a single embedding which is then fed to a classifier for all 20 instruments. The model has 226068 parameters.
\end{compactenum}

Source code for the Pytorch implementation of the neural network models is publicly available.\footnote{https://github.com/SiddGururani/AttentionMIC}
For each model, we train 10 randomly initialized instances with different random seeds and compute the classification metrics for each. This gives us a distribution of each model's performance. 
One benefit of ATT over the FC and RNN models is its small size. Both the ATT and FC\_T utilize weight-sharing for embedding instances from the bags. This leads to significantly fewer learnable parameters compared to FC and RNN while performing better than both of these models.

\subsection{Metrics}

While the total number of clips per instrument label in the OpenMIC dataset is balanced, the number of positive and negative examples is not well balanced for each instrument label. Therefore, we separately compute the precision, recall and F1-score for the positive and negative class. Thereafter, we compute the macro-average of these metrics to report the final instrument-wise metrics, meaning that positive and negative examples are weighted equally. We call these the instrument-wise precision, recall, and F1-score. Additionally, to measure the overall performance of a classifier, we macro-average the instrument-wise precision, recall and F1-score. We use a fixed threshold of $0.5$ to convert the outputs into binary predictions for computing the classification metrics.
\begin{figure}
  \centering
  \includegraphics[width=0.95\linewidth]{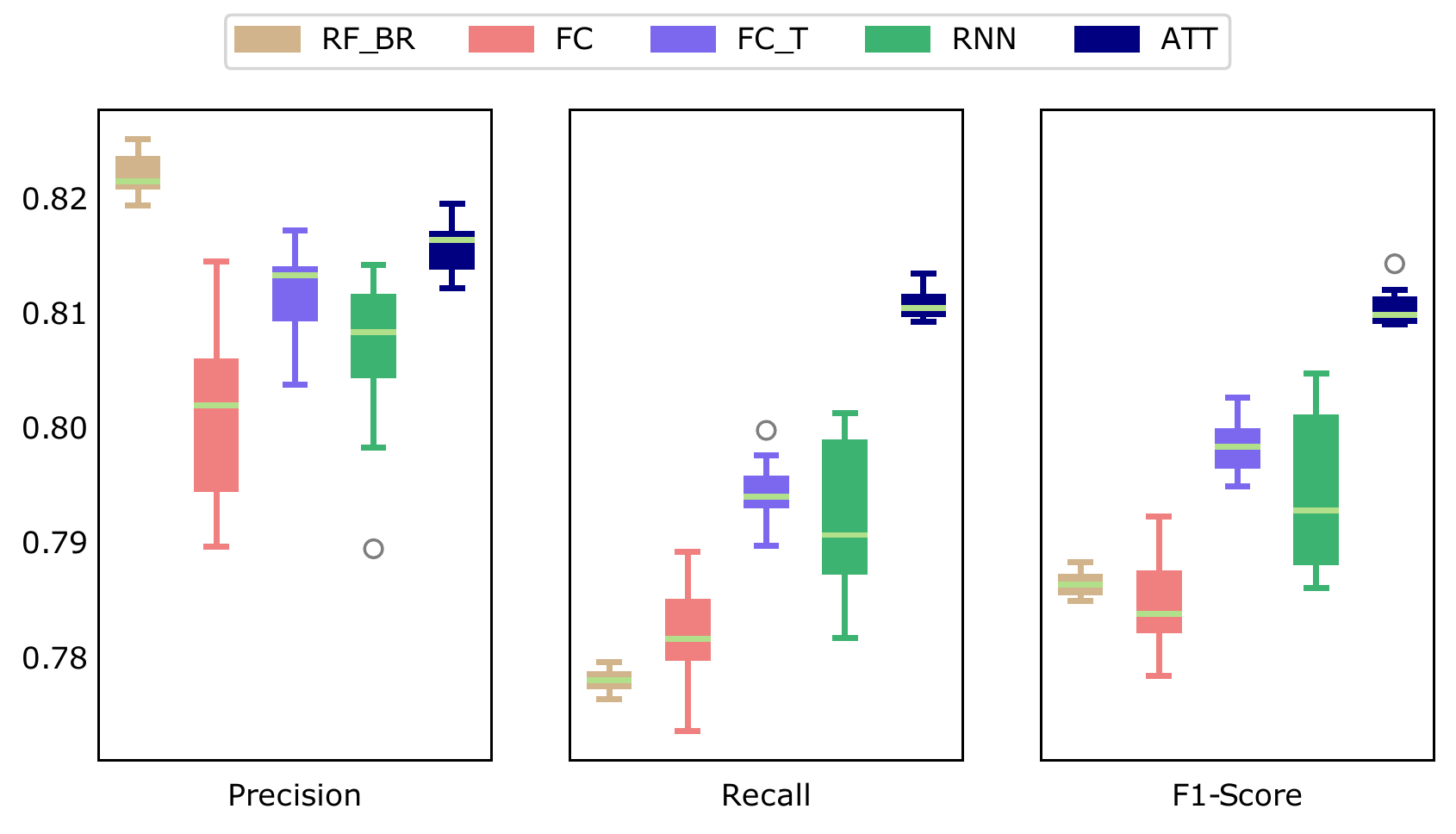}
  \caption{Precision, recall, and F1-score for different models}
  \label{fig:fig2}
\end{figure}
\section{Results and Discussion}\label{sec:discussion}

\begin{figure*}
  \centering
  \includegraphics[width=1\textwidth]{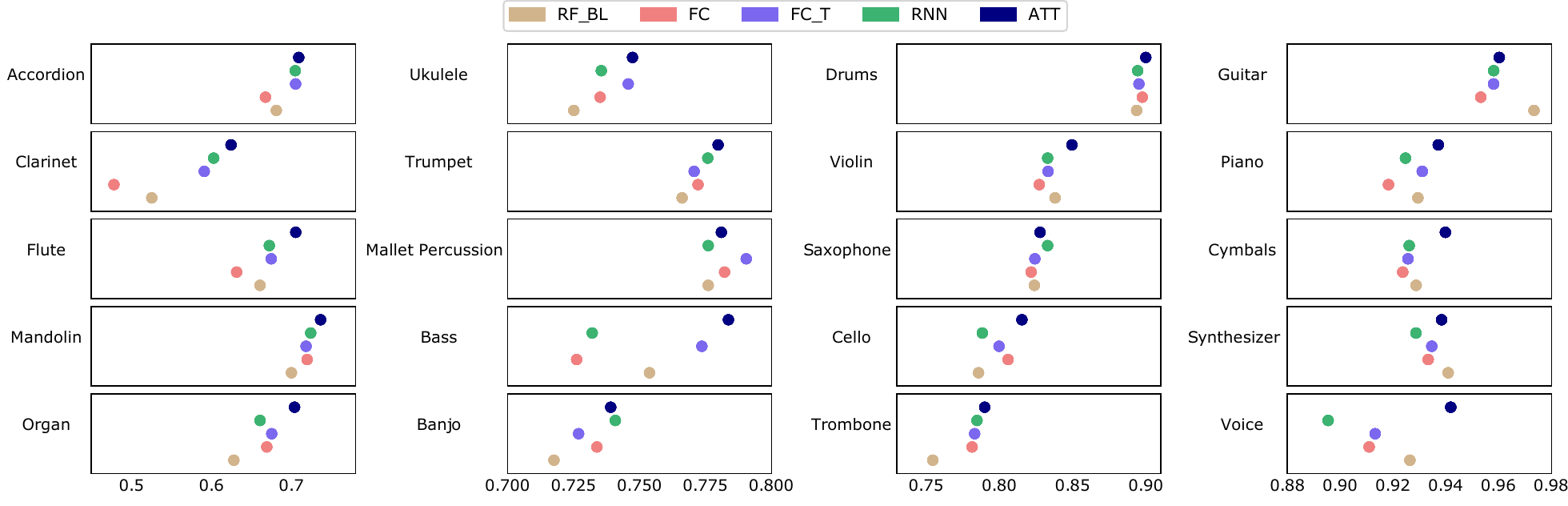}
  \caption{Instrument-wise F1-scores}
  \label{fig:fig5}
\end{figure*}

\figref{fig:fig2} shows the overall performance of ATT compared to the baseline models with box plots for the macro-averaged precision, recall, and F1-score. Additionally, we compare the instrument-wise F1-score for each model in \figref{fig:fig5}. Note that we only show the mean instrument-wise F1-score across $10$ seeds in \figref{fig:fig5} for improved visibility.

We observe that while the attention mechanism does not lead to an improvement in precision compared to the other models, the recall is improved significantly and consequently the F1-score is also improved.
We also observe that ATT performs better than RF\_BR in almost every instrument label, especially for the labels with high positive-negative class imbalance, such as clarinet, flute, and organ. This ties to the observation made about improved recall, as ATT is able to overcome this imbalance possibly due to the ability to localize the relevant instances for the minority class.
In the case of an imbalanced instrument label, the recall for the minority class greatly suffers for RF\_BR. 
While this problem is easily mitigated in standard multi-class problems by using balanced sampling, it is difficult to address with multi-label data.
Comparing to FC\_T, we can attribute the better performance of ATT to better aggregation of instance-level predictions. FC\_T is essentially the same model as ATT using mean pooling instead of attention, and ATT outperforms it for most instrument classes, especially the generally more difficult to classify instruments. 
The RNN model also beats the RF\_BR baseline. In polyphonic music, the instances in a bag are structured and highly correlated and hence using a recurrent network to model the temporal structure in the instance sequence leads to a powerful embedding of the bag, incorporating useful information from each instance.

\begin{figure}
  \centering
  \includegraphics[width=1.\linewidth]{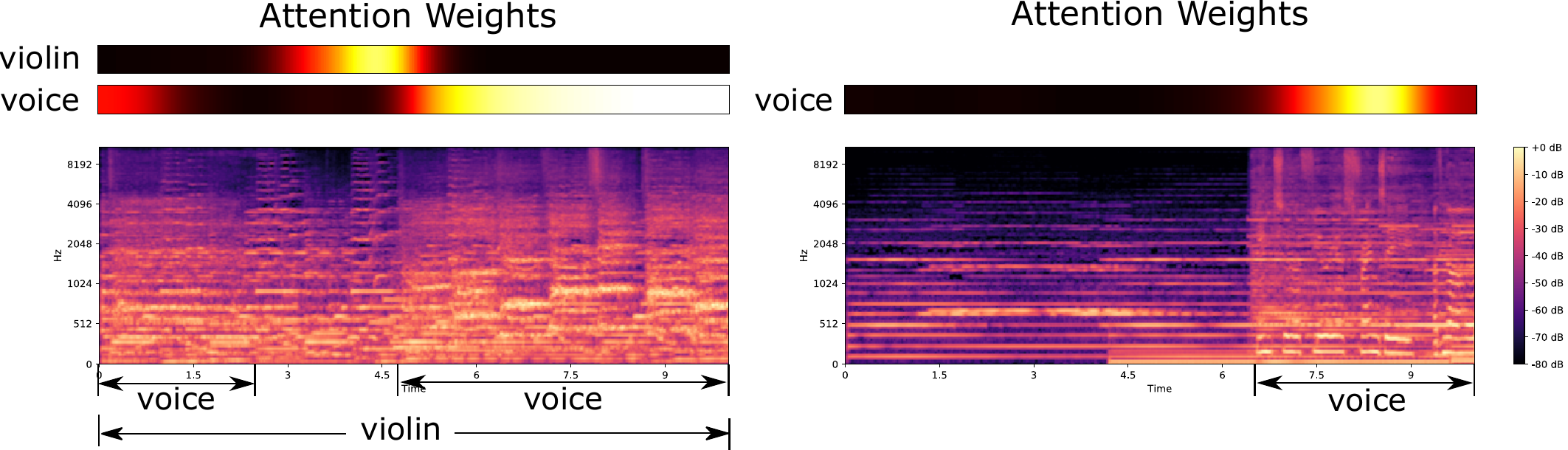}
  \caption{Attention Weight Visualization: The horizontal bars above the mel-spectrogram represent the attention weights across the instances of the clip for the respective instruments.}
  \label{fig:fig6}
\end{figure}

We visualize the attention weights for two example clips in Figure~\ref{fig:fig6}.
The left clip is from the test set and starts with the vocals fading out until 2 seconds. 
From 5 second onwards, the vocals grow in loudness until the end of the clip. 
The violin plays throughout but is the pre-dominant instrument only for a few seconds between 3 and 6 seconds, as visualized in the corresponding attention weights as well. 
The right clip is from the training set and contains vocals starting from 6 second onwards. 
The attention weights for vocals directly coincides with that. 
It is interesting to note that the annotation for vocals was missing for this clip.



\section{Conclusion}\label{sec:conclusion}

Weakly labeled datasets for instrument recognition in polyphonic music are easier to develop or annotate
 than strongly labeled datasets. This calls for a paradigm shift in the approaches towards supervised learning approaches better suited for weakly labeled data. We formulate the instrument recognition task as a MIML problem and introduce an attention-based model, evaluated on the OpenMIC dataset for 20 instruments, and compared against several other baseline models including:
\begin{inparaenum} [(i)]
  \item binary-relevance random forest, 
  \item fully connected networks, and 
  \item recurrent neural networks, 
\end{inparaenum}
We find that the attention mechanism improves the overall performance as well as the instrument-wise performance of the model while keeping the model light-weight. The example visualizations show that the model indeed is able to attend to relevant sections on a clip.

Some of the assumptions made in the formulation of the MIML problem are strong and may be worth relaxing due to the nature of musical data. We plan to further explore the task of instance-level embeddings using recurrent networks or using self-attention mechanisms as used in Transformer networks \cite{vaswani_attention_2017}. Additionally, we
 plan to address the problem of missing labels or label sparsity in the OpenMIC dataset using the curriculum learning-based methods proposed in \cite{durand_learning_2019}. Our concern is that the dataset is not large enough with enough labels for strictly supervised learning approaches to significantly improve the results much further than what we achieve with the attention mechanism, and we therefore plan to tackle the problem from other angles, such as handling missing labels or data augmentation.



\section{Acknowledgements}
This research is partially funded by Gracenote, Inc. We thank them for their generous support and meaningful discussions. 
We also thank Nvidia Corporation for their donation of a Titan V awarded as part of the GPU grant program.


\bibliography{ISMIRtemplate}

%
%
%
%

\end{document}